\documentclass[aps,prd,showpacs,preprint,preprintnumbers,nobibnotes,superscriptaddress,nofootinbib]{revtex4}
\usepackage{amsmath,amssymb,amsbsy,color}
\usepackage{epsfig}

\begin{document}
\preprint{KEK-CP-214}
\preprint{NTUTH-08-505B}
\preprint{UTHEP-566}
\preprint{YITP-08-52}

\title{Lattice study of vacuum polarization function and determination of 
strong coupling constant }

\author{E. Shintani}
\email{shintani@post.kek.jp}
\affiliation{High Energy Accelerator Research Organization (KEK), Tsukuba 305-0801, Japan}

\author{S. Aoki}
\affiliation{Graduate School of Pure and Applied Sciences, University of Tsukuba, Tsukuba 305-8571, Japan}

\author{T. W. Chiu}
\affiliation{Physics Department, Center for Theoretical Sciences, and National Center for Theoretical Sciences,
National Taiwan University, Taipei~10617, Taiwan}

\author{S. Hashimoto}
\affiliation{High Energy Accelerator Research Organization (KEK), Tsukuba 305-0801, Japan}
\affiliation{School of High Energy Accelerator Science, The Graduate University for Advanced Studies (Sokendai),
Tsukuba 305-0801, Japan}

\author{T. H. Hsieh}
\affiliation{Research Center for Applied Sciences, Academia Sinica, Taipei~115, Taiwan}

\author{T. Kaneko}
\affiliation{High Energy Accelerator Research Organization (KEK), Tsukuba 305-0801, Japan}
\affiliation{School of High Energy Accelerator Science, The Graduate University for Advanced Studies (Sokendai),
Tsukuba 305-0801, Japan}

\author{H. Matsufuru}
\affiliation{High Energy Accelerator Research Organization (KEK), Tsukuba 305-0801, Japan}

\author{J. Noaki}
\affiliation{High Energy Accelerator Research Organization (KEK), Tsukuba 305-0801, Japan}

\author{T. Onogi}
\affiliation{Yukawa Institute for Theoretical Physics, Kyoto University, Kyoto 606-8502, Japan}

\author{N. Yamada}
\affiliation{High Energy Accelerator Research Organization (KEK), Tsukuba 305-0801, Japan}
\affiliation{School of High Energy Accelerator Science, The Graduate University for Advanced Studies (Sokendai),
Tsukuba 305-0801, Japan}

\collaboration{JLQCD and TWQCD collaboration}

\begin{abstract}
  We calculate the vacuum polarization functions on the lattice using the overlap fermion formulation.
  By matching the lattice data at large momentum scales with the
  perturbative expansion supplemented by 
  Operator Product Expansion (OPE), we extract the strong coupling
  constant $\alpha_s(\mu)$ in two-flavor QCD as 
  $\Lambda^{(2)}_{\overline{MS}}$ = $0.234(9)(^{+16}_{-\ 0})$~GeV,
  where the errors are statistical and systematic, respectively.
  In addition, from the analysis of the difference between the vector
  and axial-vector channels, we obtain some of the four-quark condensates.
\end{abstract}
\maketitle

\section{Introduction}

In Quantum Chromodynamics (QCD) the vacuum polarization, 
defined through the (axial-)vector current correlator, contains rich
information of its perturbative and non-perturbative dynamics.
In the long distance regime it is sensitive to the low-lying particle spectrum.
The short distance regime, on the other hand, can be analyzed using
perturbation theory supplemented by the Operator Product Expansion (OPE).
The current correlator can be expressed as an expansion in terms of  
the strong coupling constant $\alpha_s$ together with power
corrections of the form $\langle\mathcal{O}^{(n)}\rangle/Q^n$. 
Here, the local operator $\mathcal{O}^{(n)}$ has a mass dimension $n$
and $Q$ is the momentum scale flowing into the correlator.
Determination of $\alpha_s$ 
(and of the vacuum expectation values $\langle\mathcal{O}^{(n)}\rangle$, in principle)
can be performed by applying the formulae for experimental results of
$e^+e^-$ cross section or $\tau$ decay distributions
\cite{Davier:2005xq}, for instance. 
On the other hand, if one can {\it calculate} the correlators 
non-perturbatively, theoretical determination of those fundamental
parameters is made possible. 

Lattice QCD calculation offers such a non-perturbative technique.
Two-point correlators can be calculated for space-like separations.
In this work we investigate the use of the perturbative formulae of the
correlators for the lattice data obtained in the high $Q^2$ regime.
The strong coupling constant $\alpha_s$ may then be extracted.
In such an analysis, it is essential to find the region of $Q^2$ where
the perturbative expression can be applied and at the same time the
discretization error is under control.
By inspecting the numerical data, we find that this is indeed possible
at a lattice spacing $a\simeq$ 0.12~fm if we subtract the bulk of the 
discretization effects non-perturbatively.
The remaining effect can be estimated using the perturbation theory.

The idea of analyzing the short distance regime is not new: in fact,
the analysis of hadron correlators in the whole length-scales was
proposed 15 year ago \cite{Shuryak:1993kg}, 
but to our knowledge quantitative analysis including the determination
of $\alpha_s$ and $\langle\mathcal{O}^{(n)}\rangle$ has been missing
until recently. 
(Calculation of the vacuum polarization from the vector current
correlator in lattice QCD may be found in
\cite{Blum:2002ii,Gockeler:2003cw}. 
More recently, an analysis of charmonium correlator has been published
\cite{Allison:2008xk}.)

While the vacuum polarizations $\Pi_J(Q^2)$ ($J$ denotes vector or
axial-vector channel) are ultraviolet divergent and their
precise value depends on the renormalization scheme, their derivative 
$D_J(Q^2)=-Q^2d\Pi_J(Q^2)/dQ^2$, called the Adler function
\cite{Adler:1974gd}, 
is finite and renormalization scheme independent.
Therefore, the continuum perturbative expansion of $D_J(Q^2)$ 
to order $\alpha_s^3$ \cite{Surguladze:1990tg,Gorishnii:1990vf}, can
be directly applied to the lattice data. 
At relatively low $Q^2$ region, higher order terms of OPE become relevant. 
They include the parameters describing
the gluon condensate $\langle\alpha_s G^2\rangle$ and
the quark condensate $\langle m\bar{q}q\rangle$ 
(we suppress quark flavor index assuming degenerate up and down quark masses)
at $O(1/Q^4)$, 
and four-quark condensates $\langle O_8\rangle$ and $\langle O_1\rangle$ 
at $O(1/Q^6)$ \cite{Donoghue:1999ku,Cirigliano:2003kc}.
(The explicit form of $O_8$ and $O_1$ will be given in
Section~\ref{sec:V-A}.) 

We use the lattice QCD data containing two dynamical flavors described
by the overlap fermions \cite{Aoki:2008tq}. 
The simulations are performed at lattice spacing $a$ = 0.118(2)~fm on
a $16^3\times 32$ lattice.
For the details of the simulation including the choice of the lattice
actions and parameters, we refer \cite{Aoki:2008tq}.
The physical volume is about (1.9~fm)$^3$, which is relatively small
compared to the present large scale QCD simulations.
The finite volume effect is, however, not significant for the short
distance quantities considered in this work.
The quark masses $m_q$ in this analysis are 0.015, 0.025, 0.035 and
0.050 in the lattice unit, that cover the range $[m_s/6,m_s/2]$ with
$m_s$ the physical strange quark mass.
An analysis of pion mass and decay constant is presented in
\cite{Noaki:2008iy}. 

The main advantage of this data set is that both the sea and valence
quarks preserve exact chiral and flavor symmetries by the use of the
overlap fermion formulation \cite{Neuberger:1997fp,Neuberger:1998wv}.
(Although the fermionic currents used in our calculation are not conserved at
 finite lattice spacings, it does not change the following argument
 of the operator mixing.) 
The perturbative formulae for the vacuum polarizations can therefore be
applied without any modification due to explicit violation of the
chiral symmetry. 
For instance, the scalar density operator $\bar{q}q$ to define the
quark condensate is free from the leading power divergence which
scales as $1/a^3$.
This means that a term of the form $m a^{-3}/Q^4$ is forbidden in the
OPE formula as in the continuum theory.
With the Wilson-type fermion formulation, this term may appear and has
to be identified and subtracted non-perturbatively.
With the staggered fermion formulation, there is no such problem
because of its remnant chiral symmetry, while the effect of
taste-breaking may become significant when $(aQ)^2$ becomes $O(1)$.

This paper is organized as follows.
In Section~\ref{sec:vacuum_polarization} we define the vacuum
polarization functions and explain the method to calculate them on the
lattice. 
Subtraction of lattice artifacts is discussed in some detail.
Section~\ref{sec:OPE} summarizes the perturbative formulae of OPE.
Then, in Section~\ref{sec:results} we show the results of fitting of
our data with the perturbative formulae.
Estimate of the systematic errors is also given.
Conclusions are given in Section~\ref{sec:conclusion}.

\section{Vacuum polarization function}
\label{sec:vacuum_polarization}

\subsection{Definition}
In the continuum theory, the vacuum polarization functions
$\Pi_J^{(\ell)}(Q^2)$ are defined through two-point correlation functions as
\begin{eqnarray}
  \label{eq:Jmunu}
  \lefteqn{\langle J_\mu J_\nu\rangle(Q) 
  \equiv \int d^4 x e^{iQ\cdot x}\langle 
  T\{J_\mu^{ij}(x)J_\nu^{ji}(0)\}\rangle}
  \nonumber\\
  &=& 
  (\delta_{\mu\nu}Q^2-Q_\mu Q_\nu)\Pi_J^{(1)}(Q^2)
  -Q_\mu Q_\nu \Pi_J^{(0)}(Q^2),
  \label{eq:VVcont}
\end{eqnarray}
where the current $J_\mu^{ij}$ may either be a vector current 
$V_\mu^{ij} = \bar q_i \gamma_\mu q_j$ or an axial-vector current 
$A_\mu^{ij} = \bar q_i\gamma_\mu\gamma_5 q_j$ with flavor indices $i\ne j$.
$\Pi_J^{(1)}(Q^2)$ and $\Pi_J^{(0)}(Q^2)$ denote the transverse and
longitudinal parts of the vacuum polarization, respectively.
For the vector channel ($J=V$), $\Pi_V^{(0)}(Q^2)=0$ is satisfied due to
current conservation.
For the axial-vector channel ($J=A$), the longitudinal component may
appear when the quark mass is finite.

In the lattice calculation we employ the overlap fermion formulation
\cite{Neuberger:1997fp,Neuberger:1998wv},
for which the Dirac operator is given by
\begin{equation}
  D(m) = \left(m_0+\frac{m}{2}\right)
  + \left(m_0-\frac{m}{2}\right)\gamma_5\mathrm{sgn}\left[H_W(-m_0)\right]
\end{equation}
for a bare quark mass $m$.
The kernel operator $H_W(-m_0)\equiv\gamma_5D_W(-m_0)$ is constructed
from the conventional Wilson-Dirac operator $D_W(-m_0)$ at
a large negative mass $-m_0$. 
We set $m_0=1.6$ in the numerical simulation.
We use the vector and axial-current operators of the form
\begin{eqnarray}
  \label{eq:vector}
  V_\mu^{ij} &=& Z\bar q_i \gamma_\mu\left(1-\frac{D}{2m_0}\right) q_j,
  \\
  \label{eq:axial}
  A_\mu^{ij} &=& Z\bar q_i \gamma_\mu\gamma_5 
  \left(1-\frac{D}{2m_0}\right) q_j.
\end{eqnarray}
With this choice, the vector and axial charges form a multiplet under
the axial transformation
$\delta_A^aq_i = \varepsilon \tau^a_{ij} \gamma_5(1-D/m_0)q_j$, 
$\delta_A^a\bar q_i = \varepsilon\bar q_j \tau^a_{ji} \gamma_5$, 
where $\varepsilon$ denotes an infinitesimal parameter and 
$\tau^a$ is a generator of the flavor $SU(2)$ symmetry.
The overlap fermion action is invariant under this modified chiral
transformation \cite{Luscher:1998pqa}, 
as it satisfies the Ginsparg-Wilson relation
$D\gamma_5 + \gamma_5 D = D\gamma_5 D/m_0$
\cite{Ginsparg:1981bj}.
The common renormalization factor $Z$ has been calculated
non-perturbatively as $Z$ = 1.3842(3) \cite{Noaki:2008iy}.

An obvious drawback of the (axial-)vector currents in
(\ref{eq:vector}) and (\ref{eq:axial}) is that the current
conservation property $\partial_\mu J_\mu=0$ ($J=V$ or $A$) is not
satisfied at finite lattice spacing.
It leads to a significant complication in the extraction of the
functions $\Pi_J^{(0)}(Q^2)$ and $\Pi_J^{(1)}(Q^2)$, as described in
the next subsection.
The use of the conserved (axial-)vector current \cite{Kikukawa:1998py} reduces this  
complication. Once we have extracted the functions $\Pi_J^{(0)}(Q^2)$ 
and $\Pi_J^{(1)}(Q^2)$, these two types of currents should give an  
equally good approximation to the continuum one up to the unphysical constant 
shift (and the discretization error). Our preliminary study employing  
the conserved currents shows that this is indeed the case.

\subsection{Non-perturbative subtraction of lattice artifact}

Due to the discretization effects including the current
non-conservation effect, the two-point correlation functions 
(\ref{eq:Jmunu}) may have more complicated structures.
Taking account of remaining symmetries on the lattice (parity and
cubic symmetries) but without the current conservation, the
correlators on the lattice 
$\langle J_\mu J_\nu\rangle^{\mathrm{lat}}(Q)$ can be expressed 
as an expansion in $Q_\mu$:
\begin{eqnarray}
  \lefteqn{
    \langle J_\mu J_\nu\rangle^{\rm lat}(Q)
    = \Pi_J^{(1)}(Q) {Q}^2\delta_{\mu\nu} 
    - \Pi_J^{(0+1)}(Q) {Q}_\mu{Q}_\nu
  }\nonumber\\
  &-& \sum_{n=0}^\infty B_n^J(Q) Q_\mu^{2n}\delta_{\mu\nu} 
  - \sum_{m,n=1}^\infty C_{mn}^J(Q)\big\{Q_\mu^{2m+1}Q_\nu^{2n-1}
   + Q_\nu^{2m+1} Q_\mu^{2n-1}\big\},
\label{eq:JJlat}
\end{eqnarray}
in the momentum space.
The lattice momentum ${Q}_\mu$ is defined as 
$Q_\mu = (2/a)\sin(\pi n_\mu/L_\mu)$ with an integer four-vector
$n_\mu$ whose components take values in $(-L_\mu/2,L_\mu/2]$ on a lattice of
size $L_\mu$ in the $\mu$-th direction ($L_{i=1,2,3}=16$ and $L_t=32$
in our case).
The functions corresponding to the continuum counterparts,
$\Pi_J^{(1)}(Q)$ and $\Pi_J^{(0+1)}(Q)$ 
($\equiv\Pi_J^{(0)}(Q)+\Pi_J^{(1)}(Q)$),
may also have Lorentz-violating effects and could be a function of
$Q_\mu$ in general rather than a function of just a single argument $Q^2$.

The term $B_{0}^J(Q)\delta_{\mu\nu}$, which has
the same Lorentz structure as the term of physical $\Pi_J^{(1)}(Q)$
does, contains a quadratically divergent contact term. 
Since one cannot disentangle the physical contribution from the
unphysical divergence using the Lorentz structure alone, we focus on
extracting $\Pi_J^{(0+1)}(Q)$, which is free from the contact term.

The terms including functions $B_{n>0}^J(Q)$ and $C_{mn}^J(Q)$
represent the lattice artifacts that violate the Lorentz symmetry.
They are generally written in terms of an expansion in $aQ_\mu$
and $aQ_\nu$. 
(Physically relevant terms are separately written with
a conventional notation $\Pi_J^{(1)}(Q)$ and $\Pi_J^{(0+1)}(Q)$.)
The lowest order term $B_1^J(Q)$ remains constant in the continuum
limit $aQ\to 0$, while the terms of $B_2^J(Q)$ and $C_{11}^J(Q)$ are
relatively suppressed by $O((aQ)^2)$ and vanish in the continuum
limit. 
Higher order terms are suppressed by additional powers of $a$ at a
fixed $Q$.
Since the momentum scale $Q$ of interest is not much less than the
lattice cutoff $1/a$, the convergence of the expansion at our lattice
spacing must be carefully investigated for the lattice data. 
These terms can be identified non-perturbatively, and 
we found that the lowest non-trivial terms including $B_2^J(Q)$ and
$C_{11}^J(Q)$ are already very small as described below.
Higher order terms are thus safely neglected.

Extraction of $B_{1,2}^J(Q)$ and $C_{11}^J(Q)$ from the lattice data
goes as follows.
The off-diagonal components 
$\langle J_\mu J_\nu\rangle^{\mathrm{lat}}(Q)$ ($\mu\ne\nu$)
contain $\Pi_J^{(0+1)}(Q)$ and $C_{11}^J(Q)$,
hence by taking the data with two different momentum configurations
giving the same ${Q}^2$ one can solve a linear equation to
disentangle $\Pi_J^{(0+1)}(Q)$ from the lattice artifact.
To be explicit, for two different momentum configurations $aQ^{(1)}$
and $aQ^{(2)}$ giving the same $(aQ^{(1)})^2=(aQ^{(2)})^2=(aQ)^2$, 
the linear equation is written as
\begin{eqnarray}
  \langle J_\mu J_\nu\rangle^{\rm lat}|_{\mu\ne\nu}(Q^{(1)}) 
  &=& aQ_\mu^{(1)} aQ_\nu^{(1)}\Pi_J^{(0+1)}(Q^{(1)})
  - \left(aQ^{(1)}_\mu(aQ^{(1)}_\nu)^3 +
    aQ^{(1)}_\nu(aQ^{(1)}_\mu)^3\right)
  C_{11}^J(Q^{(1)}),
  \nonumber\\
  \langle J_\mu J_\nu\rangle^{\rm lat}|_{\mu\ne\nu}(Q^{(2)}) 
  &=& aQ_\mu^{(2)} aQ_\nu^{(2)}\Pi_J^{(0+1)}(Q^{(2)})
  - \left(aQ^{(2)}_\mu(aQ^{(2)}_\nu)^3 + aQ^{(2)}_\nu(aQ^{(2)}_\mu)^3
  \right)C_{11}^J(Q^{(2)}).
  \nonumber\\
  \label{eq:JJ_linear}
\end{eqnarray}
We may assume the equalities
$\Pi_J^{(0+1)}(Q^{(1)})=\Pi_J^{(0+1)}(Q^{(2)})$ and
$C_{11}^J(Q^{(1)})=C_{11}^J(Q^{(2)})$ for small enough $(aQ)^2$,
because $aQ^{(1)}$ and $aQ^{(2)}$ are different only by permutations
of space-time directions.
The linear equation (\ref{eq:JJ_linear}) can be solved when
\begin{equation}
  \label{eq:cond}
  Q_\mu^{(1)}Q_\nu^{(1)}Q_\mu^{(2)}Q_\nu^{(2)}
  \left[ 
    (Q_\mu^{(2)})^2 + (Q_\nu^{(2)})^2 - (Q_\mu^{(1)})^2 -(Q_\nu^{(1)})^2 
  \right] \ne 0. 
\end{equation}
It is easy to see that three different non-zero components must be
contained in $aQ^{(1)}$ and $aQ^{(2)}$ to satisfy (\ref{eq:cond}).
The smallest possible momentum assignment corresponds to the
combination 
$|n_\mu^{(1)}| = (2,1,0,1)$, $|n_\mu^{(2)}| = (1,2,0,1)$ 
with $(\mu,\nu)=(1,4)$
and its permutations.
Since the fourth (temporal) direction is longer for our lattice
($L_{i=1,2,3}=16$ while $L_4=32$), the fourth component of $Q_\mu$ is
effectively 1/2 of spatial components when they are the same in
$n_\mu$. 
The corresponding momentum squared for this choice is $(aQ)^2\simeq$ 0.776.
For larger lattice momenta, there are many possible choices that this
procedure is applied.

The lattice artifact in the diagonal pieces, $B_1^J(Q)$ and
$B_2^J(Q)$, can be extracted in a similar manner by solving linear
equations for $\mu=\nu$ after subtracting the $C_{11}^J(Q)$ terms.
For instance, the leading contribution 
$(\Pi_J^{(1)}(Q)Q^2-B_0^J(Q))\delta_{\mu\nu}$
is extracted by subtracting the sub-leading contribution
$B_1^J(Q)Q_\mu^2\delta_{\mu\nu}$, which can be identified from a
difference between $\langle J_1 J_1\rangle^{\mathrm{lat}}(Q)$ and
$\langle J_2 J_2\rangle^{\mathrm{lat}}(Q)$ at the same $Q^2$, for
instance. 

\begin{figure}
  \begin{center}
    \includegraphics[width=100mm]{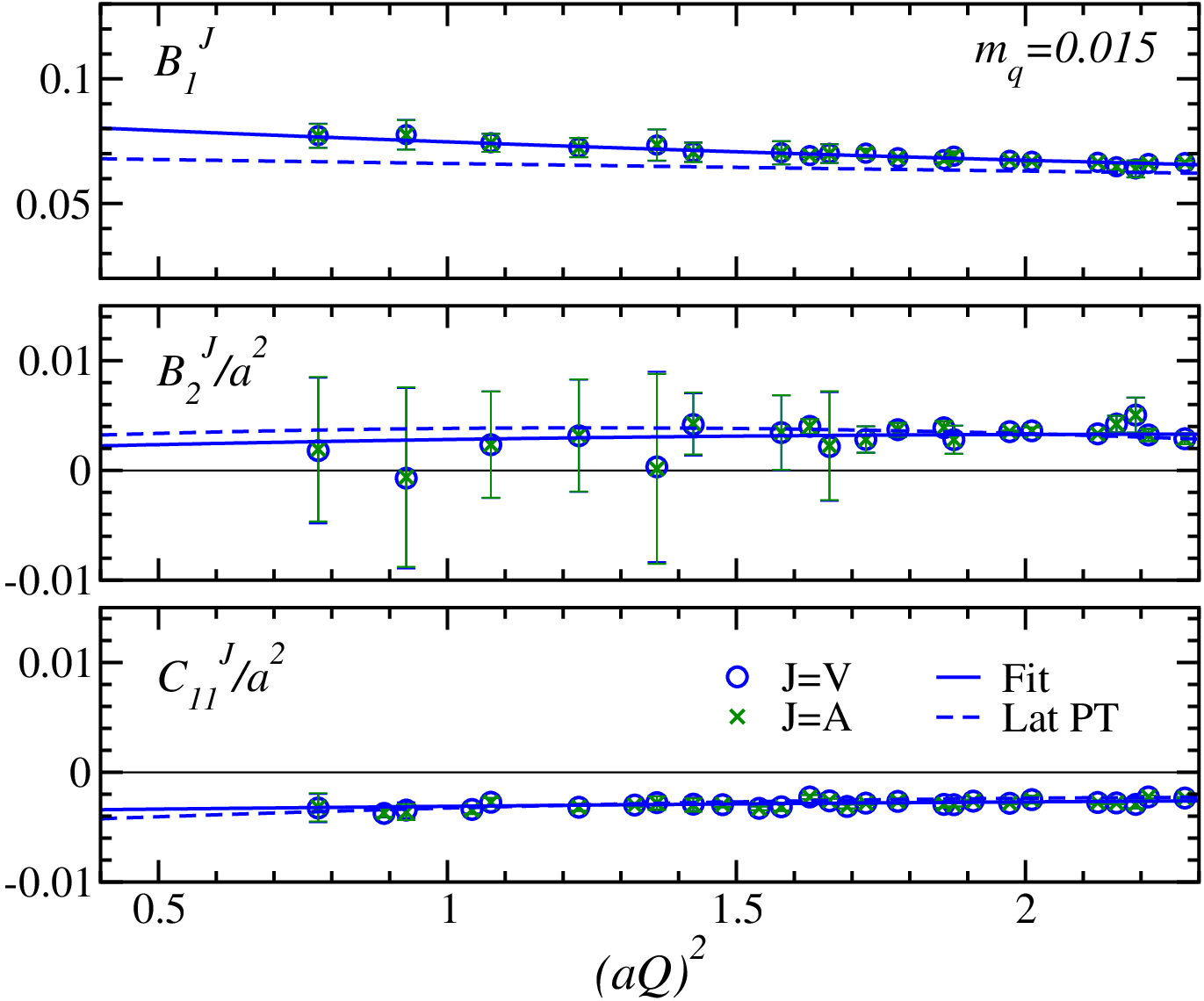}
    \caption{
      Momentum dependence of $B^J_1(Q)$, $B^J_2(Q)/a^2$, and
      $C^J_{11}(Q)/a^2$ at $m_q=0.015$. 
      Circles (crosses) show the vector (axial-vector) channel.
      The solid curves represent a polynomial fit and the dashed
      curves show the one-loop results.}
    \label{fig:BandC}
  \end{center}
\end{figure}

Figure~\ref{fig:BandC} shows the numerical results for $B_1^J(Q)$, 
$B_2^J(Q)/a^2$ and $C_{11}^J(Q)/a^2$ at the smallest quark mass ($m_q=0.015$)
as a function of $(aQ)^2$ for both vector and axial channels.
In the momentum region $(aQ)^2<$ 2.3 only the $B_1^J(Q)$ term gives
sizable contribution, while the others are an order of magnitude smaller
even without the suppression due to $(aQ)^2$.
Their dependence on $(aQ)^2$ is rather mild, so that it
seems reasonable to fit these functions as a polynomial of $(aQ)^2$.
We use a third-order polynomial to model these functions.
This is used to subtract the artifacts at the momentum points
for which the above procedure is not applicable, 
{\it e.g.} below the lowest $(aQ)^2\simeq$ 0.776.

We notice that the difference between $J=V$ and $J=A$ is consistent
with zero within statistical errors.
This indicates that these lattice artifacts are strongly constrained
by the exact chiral symmetry of the overlap fermion, and the effect of
the finite quark mass is negligible.
It also suggests that such short distance quantities are insensitive
to the spontaneous chiral symmetry breaking, as it should be.
This property is essential in the calculation of the difference
$\Pi^{(\ell)}_V(Q)-\Pi^{(\ell)}_A(Q)$, which is related to the
electromagnetic mass difference of pions 
\cite{Shintani:2007ub,Shintani:2008qe}.

\subsection{Perturbative calculation of the lattice artifacts}
\label{sec:pert}
Since the lattice artifacts are most significant in the high $(aQ)^2$
region, perturbative analysis of the discretization effects is
expected to give a reasonable estimate.
We calculate the vacuum polarization functions in the lattice
perturbation theory at one-loop level,
which means that only the zeroth order of $\alpha_s$ is included.
We then extract the terms corresponding to $\Pi_J^{(0+1)}(Q)$, 
$B_{1,2}^J(Q)$ and $C_{11}^J(Q)$. 

We calculate the vacuum polarization diagram in which two
(axial-)vector currents (\ref{eq:vector}) and (\ref{eq:axial}) are 
inserted. 
The renormalization factor $Z$ is set equal to 1 at this order.
In the momentum space, the two-point function is written as
\begin{eqnarray}
  \langle V_\mu V_\nu\rangle^{\rm lat}(Q) 
  &=& \int^\pi_{-\pi} \frac{d^4 K}{(2\pi)^4} 
  \mathrm{Tr}
  \left[ 
    \left(1-\frac{1}{2m_0}D_0(K)\right) S_0(K) \gamma_\mu
  \right. 
  \nonumber\\
  &&  
  \left.
    \times \left(1-\frac{1}{2m_0}D_0(K-Q)\right) S_0(K-Q) \gamma_\nu
  \right],
  \label{eq:LatPT}
\end{eqnarray}
where the fermion propagator $S_0(K)$ is given by
\begin{equation}
  S_0(K) =
  \frac{1}{2m_0}\left[
    \frac{-i\sum_\mu\gamma_\mu\sin(K_\mu)}{\omega(K)+b(K)}+1
  \right]
\end{equation}
with
\begin{eqnarray}
  \omega(K) & = & \sqrt{\sum_\mu \sin^2(K_\mu) + b(K)^2}, \\
  b(K) & = & \sum_\mu\left(1-\cos(K_\mu)\right)-m_0
\end{eqnarray}
for the overlap fermion and $D_0(K)^{-1}=S_0(K)$.
We set $a=1$ in this subsection.
In the perturbative calculation, $m_0$ may be set equal to 1.
At the perturbative level, the vector and axial-vector current
correlators are equivalent in the massless limit, because of the exact
chiral symmetry of the overlap fermion.

\begin{figure}
  \begin{center}
    \includegraphics[width=100mm]{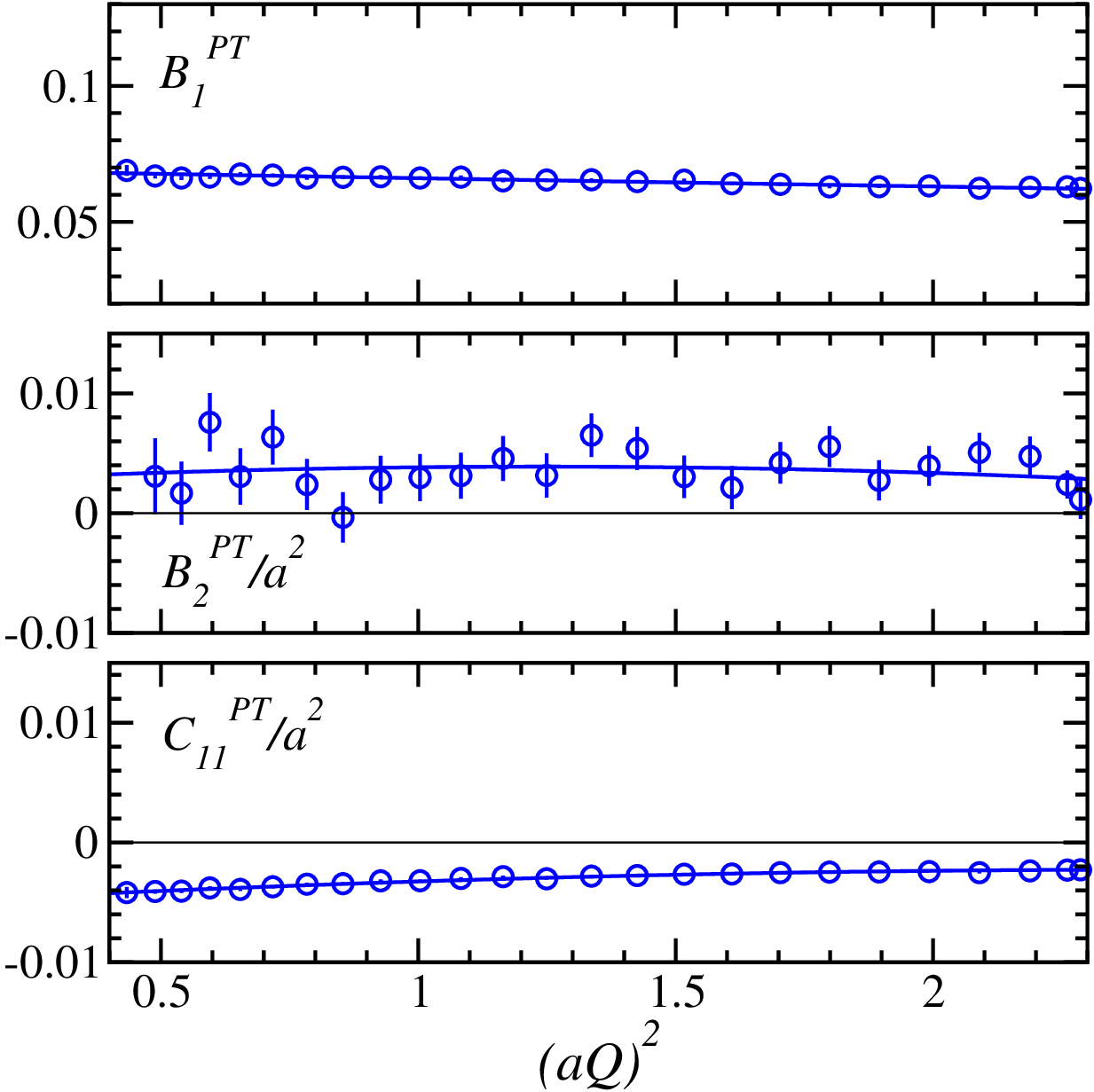}
    \caption{
      Momentum dependence of $B^J_1(Q)$, 
      $C^J_{11}(Q)/a^2$ and $B^J_2(Q)/a^2$ calculated in perturbation
      theory. 
      }
    \label{fig:BandC_PT}
  \end{center}
\end{figure}

After performing the numerical integral in (\ref{eq:LatPT}) we extract 
$B_{1,2}(Q)$, $C_{11}(Q)$ and $\Pi_V(Q)$ in (\ref{eq:JJlat})
through the same numerical procedure as we used in the
non-perturbative extraction.
To be explicit, we take representative values of $(aQ)^2$ between 0.4
and 2.3 and consider two different momentum configurations $aQ^{(1)}$
and $aQ^{(2)}$.
The results for $B_1(Q)$, $B_2(Q)/a^2$, and $C_{11}(Q)/a^2$ are shown
in Figure~\ref{fig:BandC_PT}.
As we found in the non-perturbative calculation, the $(aQ)^2$
dependence is rather mild and we may precisely model these functions by
quadratic functions:
$B_1^{PT}(Q)$ = $0.06930(59) - 0.00332(85) (aQ)^2 + 0.00009(27) (aQ)^4$,
$B_2^{PT}(Q)$ = $0.0025(22) + 0.0023(30) (aQ)^2 - 0.0009(9) (aQ)^4$,
and
$C_{11}^{PT}(Q)$ = $-0.00507(14) + 0.00227(20) (aQ)^2 - 0.00046(6) (aQ)^4$.
The fit curves are shown in Figure~\ref{fig:BandC_PT}.

The same curves are also plotted in Figure~\ref{fig:BandC} by dashed
lines. 
These perturbative results show reasonable agreement with the lattice
data. 
It indicates that the lattice artifacts are indeed well described by
the perturbation theory.

\subsection{Results for the vacuum polarization functions}

\begin{figure}
  \begin{center}
    \includegraphics[width=100mm]{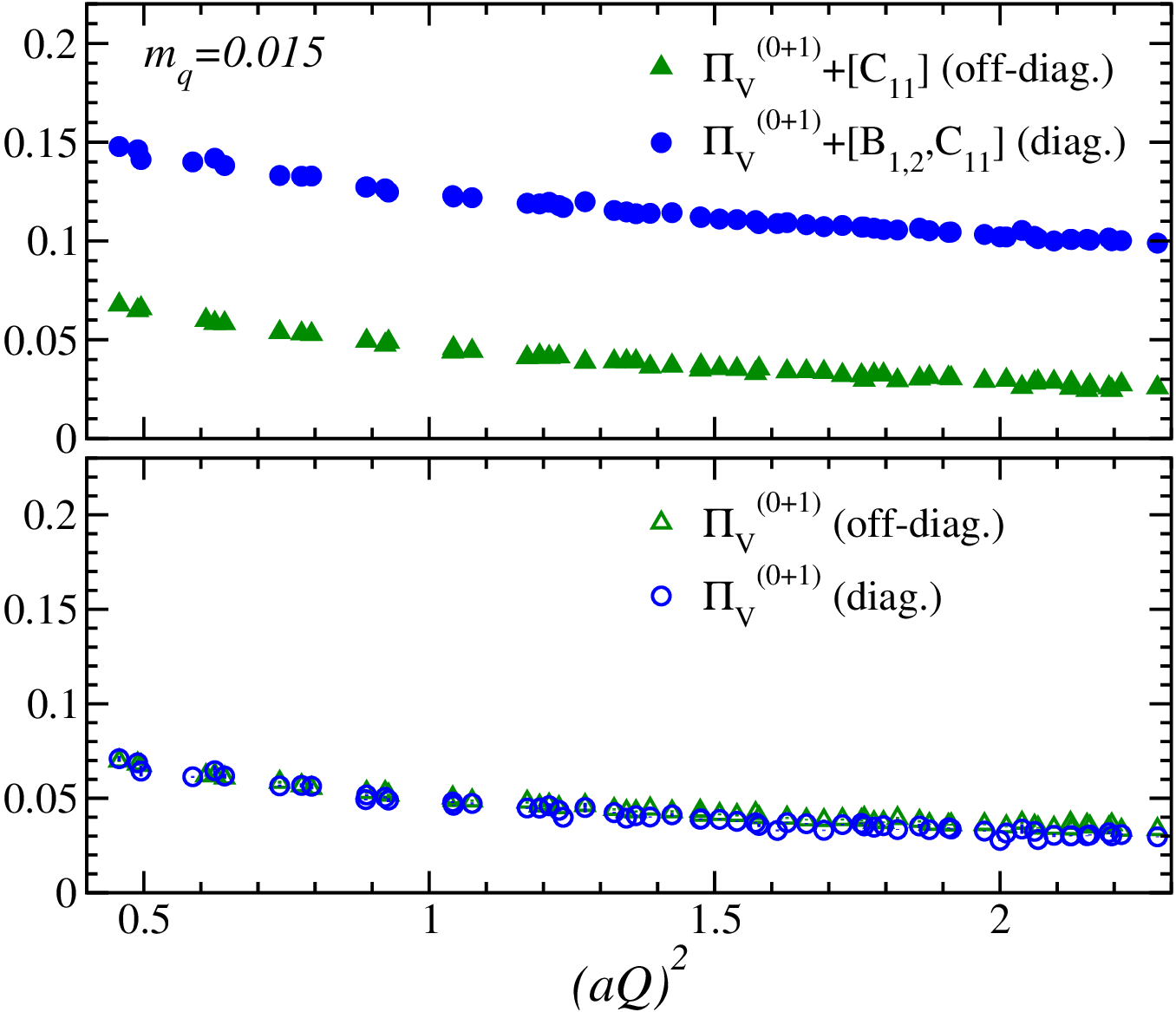}
    \caption{
      $\Pi_V^{(0+1)}(Q)$ from off-diagonal $\mu\ne\nu$ and diagonal
      $\mu=\nu$ correlators 
      with (lower panel) and without (upper panel) the
      subtraction of $B^J_{1,2}(Q)$ and $C^J_{11}(Q)$. }
    \label{fig:PiV}
\end{center}
\end{figure}

Lattice results for the vacuum polarization function
$\Pi_J^{(0+1)}(Q)$ for $J=V$ are shown in Figure~\ref{fig:PiV}.
The vacuum polarization function can be extracted 
from off-diagonal $\mu\ne\nu$ (triangles) and 
from diagonal $\mu=\nu$ (circles) components. 
Upper and lower panels show the data before and after the subtraction
of $B_n^J(Q)$ and $C_{mn}(Q)^J(Q)$ terms.
Namely, for the upper panel, $\Pi_J^{(0+1)}(Q)$ is identified with the
formula (\ref{eq:JJlat}) but without the $B_n^J(Q)$ and $C_{mn}^J(Q)$ terms.
As discussed above, raw lattice data of the diagonal components
receive large contamination from $B_1^J(Q)$ while the artifact for the
off-diagonal components is much smaller (below 0.01).

After the non-perturbative subtraction of $B_{1,2}^J(Q)$ and
$C_{11}^J(Q)$, we observe that the off-diagonal and diagonal
components give consistent results.
It strongly indicates that the higher order lattice artifacts are
unimportant.  
We average the diagonal and off-diagonal data in the following
analysis.

\section{Operator product expansion}
\label{sec:OPE}

\subsection{$V$ and $A$ channels}
We now discuss the fit of the lattice data to the
OPE expression of the form \cite{Shifman:1978bx}
\begin{eqnarray}
    \left.\Pi_J^{(0+1)}\right|_{\rm OPE}(Q^2) 
    &=& c + C_0(Q^2,\mu^2) + \frac{m^2}{Q^2}C_m^J(Q^2,\mu^2) 
    + C^J_{\bar qq}(Q^2)\frac{\langle m\bar q q\rangle}{Q^4}
    \nonumber\\
    & & + C_{GG}(Q^2)\frac{\langle(\alpha_s/\pi) GG\rangle}{Q^4}.
  \label{eq:pi_J_OPE}
\end{eqnarray}
Instead of directly treating the Adler function, we analyze its
indefinite integral $\left.\Pi_J^{(0+1)}\right|_{\rm OPE}(Q^2)$.
The coefficient functions $C_0(Q^2,\mu^2)$, $C_m^J(Q^2,\mu^2)$,
$C_{\bar qq}^J(Q^2)$ and $C_{GG}(Q^2)$ are analytically calculated in
perturbation theory.
The terms of order $1/Q^6$ and higher are not included.

A constant $c$ is divergent and thus scheme-dependent, while other
terms are finite and well-defined.
Although we need to specify the renormalization scheme, the scheme
dependence should disappear as the higher order terms are included.
The following formulae are consistently given in the
$\overline{\mathrm{MS}}$ scheme, so that the strong coupling constant 
$\alpha_s(\mu)$ is defined in this conventional scheme.

The leading term $C_0(Q^2,\mu^2)$ is known to
$\mathcal{O}(\alpha_s^2)$ in the massless limit
\cite{Surguladze:1990tg,Gorishnii:1990vf} as
\begin{eqnarray}
  C_0(Q^2,\mu^2) &=& 
  \frac{1}{16\pi^2}
  \left\{
    \frac{20}{3} + 4\ln\frac{\mu^2}{Q^2} 
    + \frac{\alpha_s(\mu^2)}{\pi}
    \left[ \frac{55}{3}-16\zeta(3) + 4\ln\frac{\mu^2}{Q^2}
    \right]
  \right.
  \nonumber\\
  & & + \left(\frac{\alpha_s(\mu^2)}{\pi}\right)^2
  \left.\left[ \frac{41927}{216} - \frac{3701}{324}N_f
    -  \left(\frac{1658}{9}-\frac{76}{9}N_f\right)\zeta(3)
    + \frac{100}{3}\zeta(5)
  \right.\right.
  \nonumber\\
  & & 
  \left.\left.
  + \left\{ \frac{365}{6}-\frac{11}{3}N_f - 
    \left(44-\frac{8}{3}N_f\right)\zeta(3) 
    +\left(\frac{11}{2}-\frac{1}{3}N_f\right)\ln\frac{\mu^2}{Q^2} 
    \right\} \ln\frac{\mu^2}{Q^2}
  \right] \right\}, \label{eq:c0}
  \nonumber\\
\end{eqnarray}
where $N_f$ denotes the number of flavors, and the zeta function is
numerically given as $\zeta(3)=1.20205\cdots$, $\zeta(5)=1.03692\cdots$.
For a finite quark mass there is a contribution of
$\mathcal{O}(m^2/Q^2)$ with running mass $m=m(\mu)$.
This term is represented by $C_m^J(Q^2,\mu^2)$, which is also calculated to
$\mathcal{O}(\alpha_s^2)$ as
\begin{eqnarray}
  C_m^V(Q^2,\mu^2) &=& \frac{1}{4\pi^2} \bigg[
  -6 + \frac{\alpha_s(\mu^2)}{\pi}\Big(-16-12\ln\frac{\mu^2}{Q^2}\Big) \nonumber\\
  &+& \Big(\frac{\alpha_s(\mu^2)}{\pi}\Big)^2 \Big\{
  -\frac{19691}{72} + \frac{95}{12}N_f - \frac{124}{9}\zeta(3) + \frac{1045}{9}\zeta(5)\nonumber\\
  &-& \Big(55 + 12\ln\frac{\mu^2}{Q^2}\Big)\ln\frac{\mu^2}{Q^2}
  - \Big(11-\frac{2}{3}N_f\Big)\Big(\frac{13}{2}+\frac{3}{2}\ln\frac{\mu^2}{Q^2}\Big)
  \ln\frac{\mu^2}{Q^2}\Big\} \bigg]\nonumber\\
  &+& \frac{N_f}{16\pi^2}\Big(\frac{\alpha_s(\mu^2)}{\pi}\Big)^2
  \Big[\frac{128}{3}-32\zeta(3)\Big],
  \\
  C_m^A(Q^2,\mu^2) &=& \frac{1}{4\pi^2} \bigg[
  -6+\frac{\alpha_s(\mu^2)}{\pi}\Big(-12-12\ln\frac{\mu^2}{Q^2}\Big) \nonumber\\
  &+& \Big(\frac{\alpha_s(\mu^2)}{\pi}\Big)^2 \Big\{ 
  -\frac{4681}{24} + \frac{55}{12}N_f - \Big(34-\frac{8}{3}N_f\Big)\zeta(3) + 115\zeta(5)
  \nonumber\\
  &-& \Big( 47+12\ln\frac{\mu^2}{Q^2}\Big)\ln\frac{\mu^2}{Q^2}-\Big(11-\frac{2}{3}N_f\Big)
  \Big(\frac{11}{2}+\frac{3}{2}\ln\frac{\mu^2}{Q^2}\Big)\ln\frac{\mu^2}{Q^2}\Big\} \bigg]
  \nonumber\\
  &+& \frac{N_f}{16\pi^2}\Big(\frac{\alpha_s(\mu^2)}{\pi}\Big)^2\Big[\frac{128}{3}-32\zeta(3)\Big].
\end{eqnarray}
We ignore terms of $\mathcal{O}(m^4)$ and higher.

The OPE corrections of the form $\langle O^{(n)}\rangle/Q^n$ start
from the dimension-four operators 
$m\bar{q}q$ and $(\alpha_s/\pi)GG$.
Their Wilson coefficients $C^J_{\bar{q}q}(Q^2)$ and
$C_{GG}(Q^2)$ are known to $\mathcal{O}(\alpha_s^2)$ and to
$\mathcal{O}(\alpha_s)$, respectively, as
\cite{Chetyrkin:1985kn}
\begin{eqnarray}
C_{\bar qq}^{V/A}(Q^2) &=& -2\frac{\alpha_s(\mu^2)}{\pi}\Big[ 1+\frac{1}{24}
   \frac{\alpha_s(\mu^2)}{\pi}\Big\{(116-4N_f)+(66-4N_f)\ln\frac{\mu^2}{Q^2}\Big\}\Big]\nonumber\\
  &+/-& 2\Big[ 1+\frac{4}{3}\frac{\alpha_s(\mu^2)}{\pi}
      + \frac{4}{3}\Big(\frac{\alpha_s(\mu^2)}{\pi}\Big)^2 \Big\{ 
         \Big(\frac{191}{24}-\frac{7}{36}N_f\Big)
        +\Big(\frac{11}{4}-\frac{1}{6}N_f\Big)\ln\frac{\mu^2}{Q^2}\Big\}\Big]
  \nonumber\\
  &+& \frac{N_f}{3}\Big(\frac{\alpha_s(\mu^2)}{\pi}\Big)^2\Big(4\zeta(3)-3+\ln\frac{\mu^2}{Q^2}\Big) + 0/4,\\
C_{GG}(Q^2) &=& \frac{1}{12}\Big[ 1 - \frac{11}{18}\frac{\alpha_s(Q)}{\pi}\Big].
\end{eqnarray}
Here we note that the ``gluon condensate''
$\langle(\alpha_s/\pi)GG\rangle$ is defined only through the 
perturbative expression like (\ref{eq:pi_J_OPE}).
Due to an operator mixing with the identity operator, the operator
$(\alpha_s/\pi)GG$ contains a quartic power divergence that cannot be
unambiguously subtracted within perturbation theory, which is known as
the renormalon ambiguity \cite{Martinelli:1996pk}.
Therefore, the term $\langle(\alpha_s/\pi)GG\rangle$ in
(\ref{eq:pi_J_OPE}) only has a meaning of a parameter in OPE, that may
depend on the order of the perturbative expansion, for instance.

The quark condensate $\langle \bar{q}q\rangle$ is, on the other hand,
well-defined in the massless limit, since it does not mix with lower
dimensional operators, provided that the chiral symmetry is preserved
on the lattice.
Power divergence may appear at finite quark mass as $ma^{-2}$.
In the OPE formula (\ref{eq:pi_J_OPE}), it thus leads to a functional
dependence $m^2a^{-2}/Q^4$.
Since the quark mass in the lattice unit is small (0.015--0.050) and
$(aQ)^2$ is of $O(1)$ in our lattice setup, this divergent
contribution is tiny ($\sim$ 0.1--0.2\%). In fact, we do not find
any significant $m^2$ dependence in the lattice data. We therefore
neglect this $m^2$ dependence in the numerical analysis.

\subsection{$V-A$ channel}
\label{sec:V-A}
In addition to the individual vector and axial-vector correlators, we
consider the $V-A$ vacuum polarization function.
For the difference $\Pi^{(0+1)}_{V-A}(Q)\equiv\Pi^{(0+1)}_V(Q)-\Pi^{(0+1)}_A(Q)$,
the lattice data are more precise than the individual
$\Pi^{(0+1)}_J(Q)$, so that the $1/Q^6$ and $1/Q^8$ terms are also necessary:
\begin{eqnarray}
  \left.\Pi_{V-A}^{(0+1)}\right|_{\rm OPE}(Q^2) 
  &=& (C_m^V-C_m^A)(Q^2)\frac{1}{Q^2}
  + \left(C^V_{\bar qq}-C^A_{\bar qq}\right)(Q^2)
  \frac{\langle m\bar q q\rangle}{Q^4}\nonumber\\
  & + & \left(a_6(\mu) + b_6(\mu)\ln\frac{Q^2}{\mu^2} + c_6 m_q\right)\frac{1}{Q^6}
  + \frac{a_8}{Q^8}.
  \label{eq:pi_V-A_OPE}
\end{eqnarray}
In the $V-A$ combination the coefficients $C_m^V-C_m^A$ and 
$C^V_{\bar qq}-C^A_{\bar qq}$ start at $\mathcal O(\alpha_s)$.
The coefficients $a_6(\mu)$ and $b_6(\mu)$ contain dimension six
operators $O_8$ and $O_1$ as \cite{Donoghue:1999ku,Cirigliano:2003kc}
\begin{eqnarray}
  a_6(\mu) &=& 2\pi\langle\alpha_sO_8\rangle(\mu)
  + \frac 25 4\langle\alpha_s^2 O_8\rangle(\mu) 
  + 2\langle \alpha_s^2 O_1\rangle(\mu),\\
  b_6(\mu) &=& -\langle\alpha_s^2 O_8\rangle(\mu) 
  + \frac 8 3\langle \alpha_s^2 O_1\rangle(\mu),
\end{eqnarray}
and the definition of these operators is given by
\begin{eqnarray}
  \langle O_8\rangle &=& \sum_{\mu,i,j} \left\langle 
    (\bar q_i\gamma_\mu\tau^3_{ij} q_j)(\bar q_i\gamma_\mu\tau^3_{ij} q_j)
    -(\bar q_i\gamma_\mu\gamma_5\tau^3_{ij} q_j)
    (\bar q_i\gamma_\mu\gamma_5\tau^3_{ij} q_j) \right\rangle,\\
  \langle O_1\rangle &=& \sum_{\mu,a,i,j} \left\langle 
    (\bar q_i\gamma_\mu\lambda^a\tau^3_{ij} q_j)
    (\bar q_i\gamma_\mu\lambda^a\tau^3_{ij} q_j)
   -(\bar q_i\gamma_\mu\gamma_5\lambda^a\tau^3_{ij} q_j)
    (\bar q_i\gamma_\mu\gamma_5\lambda^a\tau^3_{ij} q_j) \right\rangle,
\end{eqnarray}
with generator matrices $\tau^3$ and $\lambda^a$ of flavor SU(2) and
color SU(3) symmetries, respectively.
The numerical coefficients in the definition of $a_6$ and $b_6$
correspond to those of the Naive Dimensional Regularization (NDR) of
$\gamma_5$.  

Unlike the dimension-four quark condensate $\langle m\bar{q}q\rangle$,
$\langle O_8\rangle$ and $\langle O_1\rangle$ remain finite in the
massless limit, hence gives leading contribution.
The term $c_6$, which has a mass-dimension five, 
describes their dependence on the quark mass.
The term $a_8/Q^8$ represents the contributions from dimension-eight
operators.

\section{Fitting results}
\label{sec:results}

\subsection{Fit parameters}
In the fitting of the lattice data with the functions
(\ref{eq:pi_J_OPE}) and (\ref{eq:pi_V-A_OPE}), we fix the scale $\mu$
to 2~GeV. 
We use the value of the quark condensate obtained from a simulation in
the $\epsilon$-regime using the same lattice formulation at slightly
smaller lattice spacing, 
$\langle\bar{q}q\rangle(\mathrm{2~GeV})$ = $-$[0.251(7)(11)~GeV]$^3$
\cite{Fukaya:2007fb}.
(The values quoted in \cite{Fukaya:2007yv,Fukaya:2007pn,Noaki:2008iy}
are slightly different from but consistent with this number.
The precise value does not affect the fit much, since the contribution of
the $C_{\bar qq}$ term is sub-dominant.)
The quark mass is renormalized in the $\overline{\mathrm{MS}}$ scheme
using the non-perturbative matching factor $Z_m(\mathrm{2~GeV})$ =
0.838(17) \cite{Noaki:2008iy} as $m(\mu)=Z_m(\mu)m_q$.
The coupling constant $\alpha_s(\mu)$ is transformed to the scale of
two-flavor QCD, $\Lambda^{(2)}_{\overline{\mathrm{MS}}}$,
using the four-loop formula \cite{van Ritbergen:1997va}
\begin{eqnarray}
  \frac{\alpha_s(\mu^2)}{\pi} 
  &=& \frac{1}{\beta_0 L}
  \left[ 
    1-\frac{\beta_1}{\beta_0^2}\frac{\ln L}{L}
    + \frac{1}{\beta_0^2L^2}
    \left\{\frac{\beta_1^2}{\beta_0^2}
      \left(\ln^2L-\ln L-1\right)+\frac{\beta_2}{\beta_0}
    \right\}
  \right.
  \nonumber\\
  & &
  \left. +
    \frac{1}{\beta_0^3L^3}\left\{\frac{\beta_1^3}{\beta_0^3}
      \left(-\ln^3L+\frac{5}{2}\ln^2L+2\ln L-\frac{1}{2}\right)
      -3\frac{\beta_1\beta_2}{\beta_0^2}\ln L+\frac{\beta_3}{2\beta_0}
    \right\}
  \right]
\end{eqnarray}
with
\begin{eqnarray}
  \beta_0 &=& \frac{1}{4}\left(11-\frac{2}{3}N_f\right),\\
  \beta_1 &=& \frac{1}{4^2}\left(102-\frac{38}{3}N_f\right),\\
  \beta_2 &=& \frac{1}{4^3}\left(\frac{2857}{2}-\frac{5033}{18}N_f + 
    \frac{325}{54}N_f^2\right),\\
  \beta_3 &=& \frac{1}{4^4}\left[\frac{149753}{6}+3564\zeta(3)-
    \left(\frac{1078361}{162}+\frac{6508}{27}\zeta(3)\right)N_f
    \right.
    \nonumber\\
    && +
    \left. \left(\frac{50065}{162}+\frac{6472}{81}\zeta(3)\right)N_f^2+
      \frac{1093}{729}N_f^3\right],
\end{eqnarray}
and $L=\ln(\mu^2/\Lambda_{\overline{MS}}^{(N_f)2})$.

Then, the free parameters in the fit are the scheme-dependent constant
$c$, the gluon condensate paemeter $\langle(\alpha_s/\pi)GG\rangle$, and 
the QCD scale $\Lambda^{(2)}_{\overline{\mathrm{MS}}}$ for the fit of an average
$\Pi_{V+A}^{(0+1)}(Q)\equiv\Pi_V^{(0+1)}(Q)+\Pi_A^{(0+1)}(Q)$.
For the difference $\Pi_{V-A}^{(0+1)}(Q)$, 
$\Lambda^{(2)}_{\overline{\mathrm{MS}}}$ obtained above is used as an input
and the dimension-six condensates $a_6$, $b_6$ and $c_6$ are free
parameters.

\subsection{$V+A$ channel}

\begin{figure}
  \begin{center}
    \includegraphics[width=100mm]{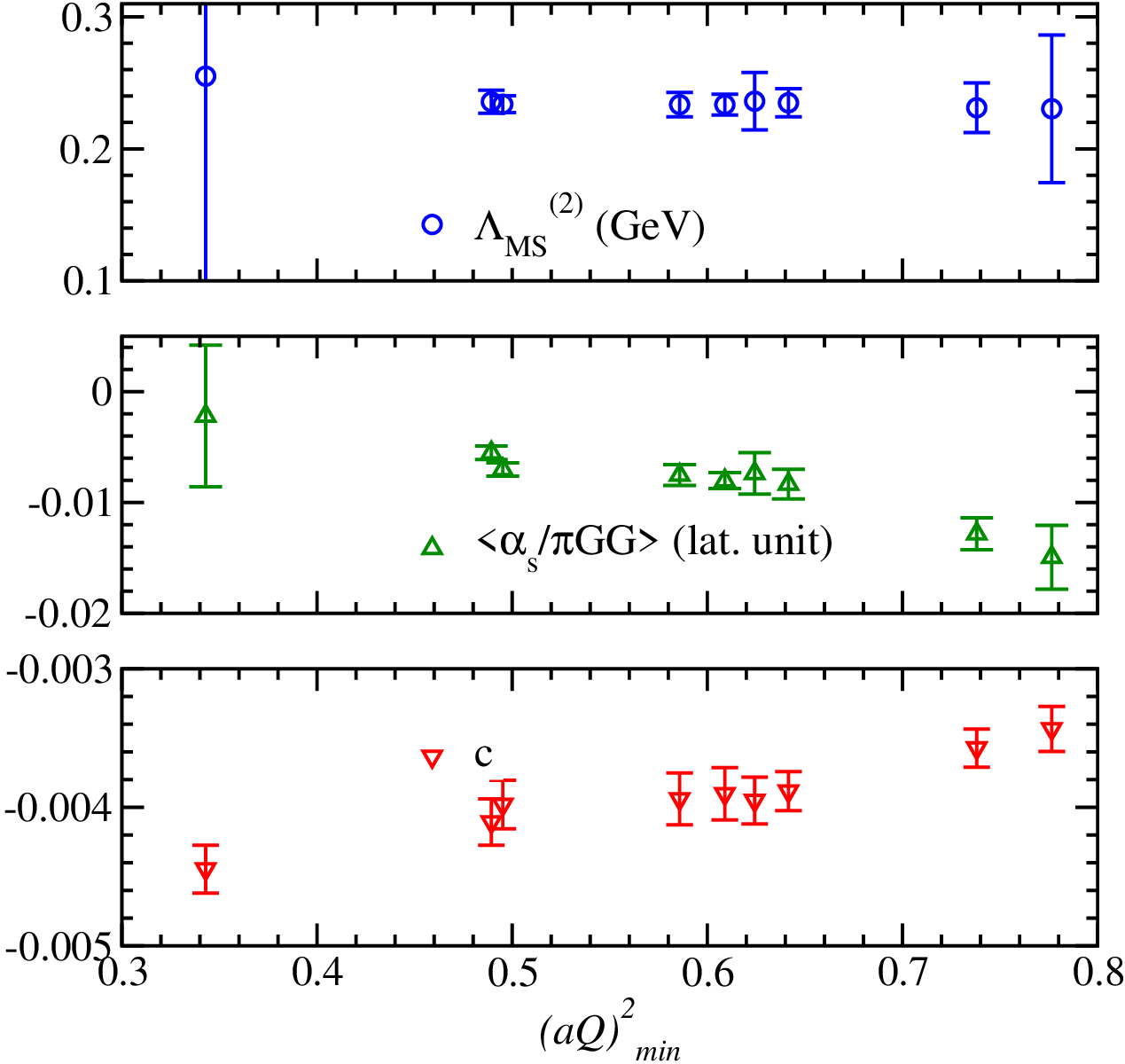}
    \caption{
      Fit range dependence of $\Lambda^{(2)}_{\overline{MS}}$, 
      $\langle(\alpha_s/\pi)GG\rangle$ and the constant term $c$.
      The maximum momentum squared $(aQ)^2_{max}$ is 1.324.
    }
    \label{fig2a}
  \end{center}
\end{figure}

The OPE analysis requires a window in $Q^2$ where the systematic errors are under control. 
The upper limit $(aQ)_{\mathrm{max}}^2\simeq 1.3238$ is set by taking the points
where different definitions of the lattice momentum, {\it i.e.} 
$Q_\mu = (2/a)\sin(\pi n_\mu/L_\mu)$ and 
$Q_\mu=(2/a)\pi n_\mu/L_\mu$, give consistent results within one
standard deviation.
In the physical unit, this corresponds to 1.92~GeV.
To determine $(aQ)_{\mathrm{min}}^2$, we investigate the dependence of
the fit parameters on $(aQ)_{\mathrm{min}}^2$ in Figure~\ref{fig2a}.
We observe that the results for
$\Lambda^{(2)}_{\overline {\mathrm{MS}}}$, 
$\langle(\alpha_s/\pi)GG\rangle$, and $c$ are stable between
$(aQ)_{\mathrm{min}}^2\simeq$ 0.48 and 0.65, which correspond to the
momentum scale 1.16--1.35~GeV.
Above $(aQ)_{\mathrm{min}}^2\simeq 0.65$ the fit becomes unstable;
the results are still consistent within one standard deviation.

\begin{figure}
  \begin{center}
    \includegraphics[width=100mm]{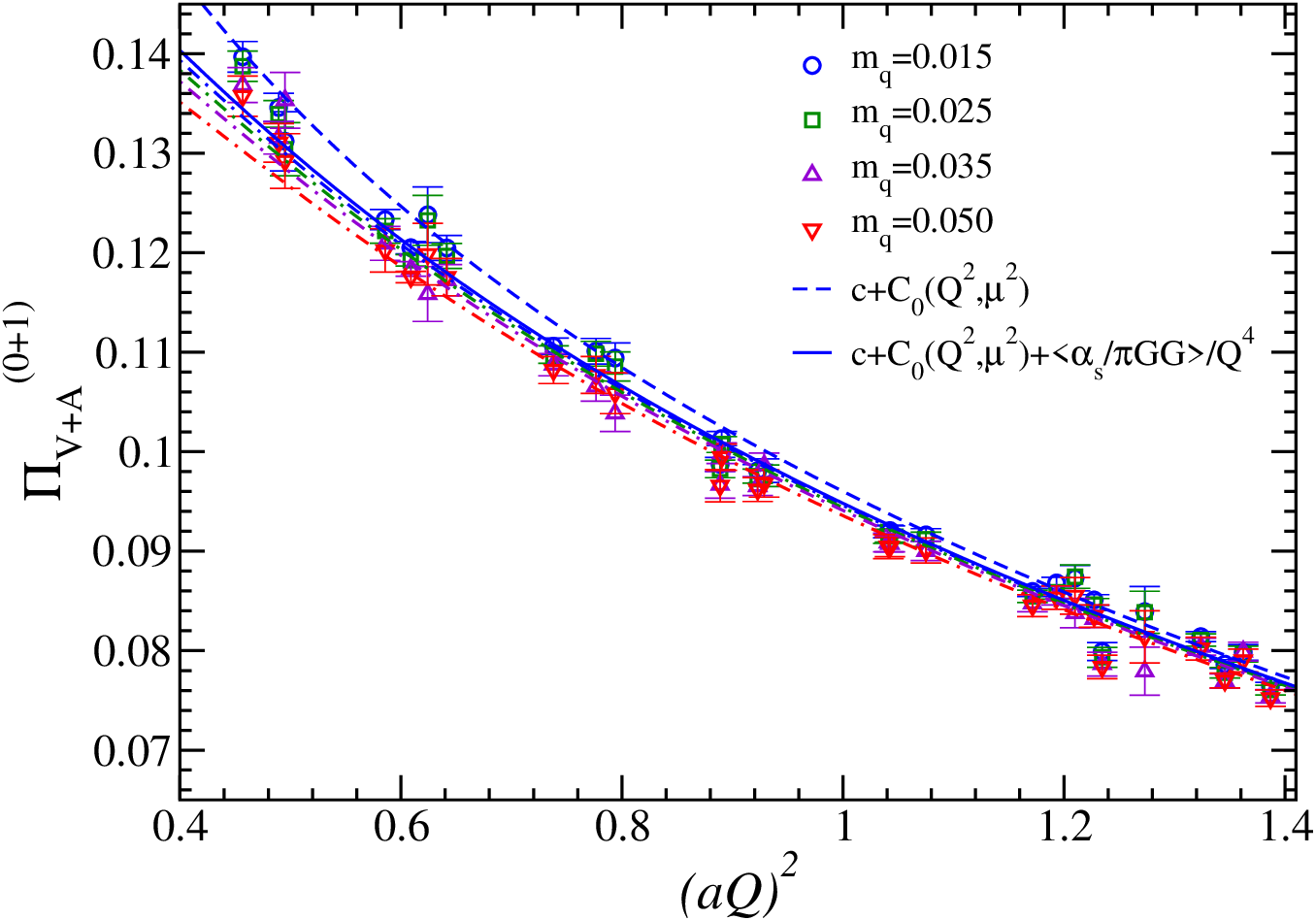}
    \caption{
      $\Pi^{(0+1)}_{V+A}(Q)$ as a function of $(aQ)^2$.
      The lattice data at different quark masses are shown by open symbols.
      Fit curves for each quark mass and in the chiral limit are drawn. 
      The full result in the chiral
      limit (dashed-dots curves are at the finite masses, and  solid curve is in the chiral limit), 
      as well as that without $\langle\alpha_s G^2\rangle/Q^4$ term (dashed curve), are shown.
    }
    \label{fig2b}
  \end{center}
\end{figure}

Figure~\ref{fig2b} shows the lattice data for
$\Pi_{V+A}^{(0+1)}(Q)$ at each quark mass and corresponding fit
curves.
It is clear that the $Q^2$ dependence of the lattice data is well
reproduced by the analytic formula.
The quark mass dependence of $\Pi_{V+A}^{(0+1)}(Q)$ is, on the other
hand, not substantial as expected from the fit function
(\ref{eq:pi_J_OPE}).  
Our fit with the known value of $\langle\bar{q}q\rangle$ reproduces
the data well.
In the chiral limit, (\ref{eq:pi_J_OPE}) is controlled by two
parameters: $\Lambda^{(2)}_{\overline{MS}}$ and 
$\langle(\alpha_s/\pi)GG\rangle$ (apart from the unphysical constant
term $c$). 
The fit result in the chiral limit is drawn by a solid curve.
The dashed curve drifting upwards towards low $Q^2$ region shows the
result when the contribution from the $\langle(\alpha_s/\pi)GG\rangle$
term is omitted by hand.
It indicates that the $Q^2$ dependence is mainly controlled by
the perturbative piece while the dimension-four term gives a minor
contribution, which becomes slightly more important in the low $Q^2$
regime.
Numerically, we obtain $\Lambda^{(2)}_{\overline{MS}}$ = $0.234(9)$~GeV and
$\langle(\alpha_s/\pi)GG\rangle$ = $-0.058(7)$~GeV$^4$
from a global fit of the lattice data at four different quark masses.
The fit range of $(aQ)^2$ is [0.65, 1.3238].

\begin{figure}
  \begin{center}
    \includegraphics[width=100mm]{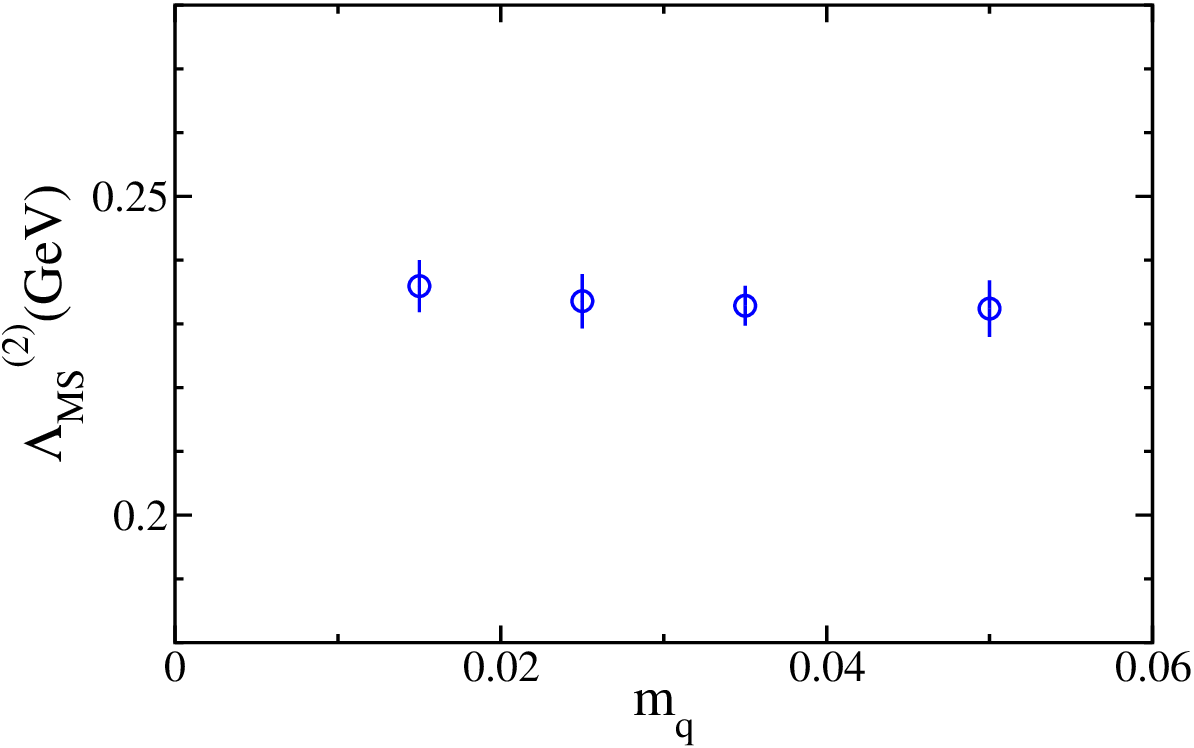}
    \caption{
      $\Lambda^{(2)}_{\overline{MS}}$ from the data at each quark mass.
    }
    \label{fig3}
  \end{center}
\end{figure}

Figure~\ref{fig3} shows $\Lambda^{(2)}_{\overline{MS}}$ 
extracted from the lattice data at each quark mass.
The flat behavior provides another evidence that the lattice data are
consistent with the perturbative prediction (\ref{eq:pi_J_OPE}).

\subsection{Systematic errors}
In this subsection we discuss on possible systematic errors in this
determination of $\Lambda^{(2)}_{\overline{MS}}$.
That includes an estimate of the discretization effects and that of
the truncation of perturbative and operator product expansions.

As indicated from the perturbative analysis presented in
Section~\ref{sec:pert}, the discretization effects are estimated
reasonably well using the perturbation theory.
Here, we discuss on the one-loop results for $\Pi_J(Q)$ on
the lattice for our choice of the fermion action and the current
operators. 
This aims at estimating the remaining systematic errors due to the
discretization effects after explicitly subtracting the $B_n^J(Q)$ and
$C_{mn}^J(Q)$ terms.

We again calculate the same one-loop vacuum polarization diagram at
representative values of $(aQ)^2$ between 0.1 and 2.0.
After subtracting the $B_{1,2}^J(Q)$ and $C_{11}^J(Q)$ terms
determined perturbatively in Section~\ref{sec:pert} we numerically
obtain the piece corresponding to $\Pi_J(Q)$, which contains the
physical logarithmic dependence $-1/(4\pi)^2\ln((aQ^2))$ as well as
the lattice artifacts. 
In the continuum theory (or the perturbative calculation with the
dimensional regularization, to be specific) only this
logarithmic term appears, hence we may identify the remaining terms as
the lattice artifacts.
They are parametrized by a polynomial of $(aQ)^2$.

\begin{figure}
  \begin{center}
    \includegraphics[width=100mm]{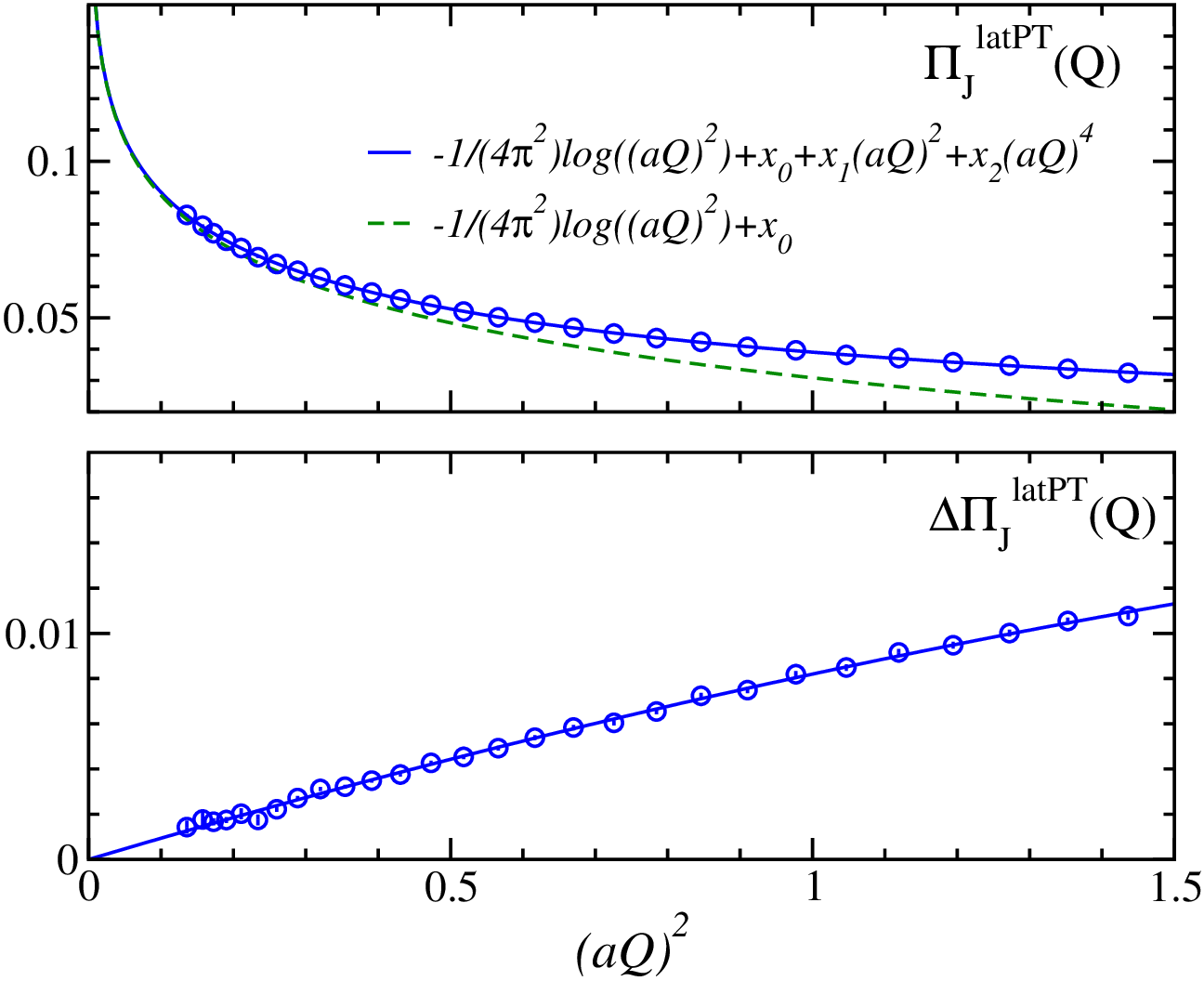}
    \caption{
      One-loop calculation of $\Pi_V(Q)$ (upper panel) at
      representative values of $(aQ)^2$ (circles) and a fit with the
      log-plus-polynomial form.
      The dashed curve shows the purely logarithmic contribution.
      The term that represents the lattice artifact $\Delta\Pi_V(Q)$
      is shown in the lower panel.
    }
    \label{fig4}
  \end{center}
\end{figure}

The result of the one-loop calculation is shown in Figure~\ref{fig4}
(upper panel). 
We fit the data with a function including the known logarithmic term
plus a quadratic function of $(aQ)^2$ and obtain the numerical result
$\Pi_V^{\rm LatPT}(Q^2) = -\frac{1}{4\pi^2}\ln((aQ)^2) 
+ 0.03085(9) + 0.00952(30)(aQ)^2 -0.00132(20)(aQ)^4$.

In order to estimate the impact of this size of the discretization
effect, we add this term to the fit function (\ref{eq:pi_J_OPE}) and
repeat the whole analysis.
The result is
$\Lambda^{(2)}_{\overline{MS}}$ = $0.249(37)$~GeV and 
$\langle(\alpha_s/\pi)GG\rangle$ = $+0.11(15)$~GeV$^4$.
We find that $\Lambda^{(2)}_{\overline{MS}}$ is not largely affected,
while $\langle(\alpha_s/\pi)GG\rangle$ is very sensitive to the
lattice artifact and in fact changes its sign.

Other (Lorentz-violating) discretization effects due to $B_n^J(Q)$ and
$C_{mn}^J(Q)$ are subtracted non-perturbatively so that the associated
error should be negligible.
With our preliminary calculation of the above mentioned conserved
vector and axial-vector currents for the overlap fermion, we confirmed
that the results are consistent with the calculation presented in this
paper obtained with the non-conserved currents (\ref{eq:vector}) and
(\ref{eq:axial}) up to the unphysical constant term $c$.
This observation confirms that our procedure to subtract the
$B_n^J(Q)$ and $C_{mn}^J(Q)$ terms is working as expected.

The truncation of the perturbative and operator product expansions
is also a possible source of the systematic error.
In order to estimate the size of the former, we repeat the analysis
using the fit formulae truncated at a lower order (two-loop level),
and find that the change of $\Lambda^{(2)}_{\overline{MS}}$ is
much less than one standard deviation.
It indicates that the higher order effects are negligible. 
The error from the truncation of OPE is estimated by dropping the
terms of ${\cal O}(1/Q^4)$ from (\ref{eq:pi_J_OPE}).
From fits with higher $(aQ)^2_{\mathrm{min}}$ (between 0.79 and 0.89)
to avoid contamination from the $1/Q^4$ effects, we obtain
$\Lambda^{(2)}_{\overline{MS}}$ = 0.247(3)~GeV.
The deviation of $\Lambda^{(2)}_{\overline{MS}}$ is about the same
size as that due to the discretization effect.

The errors due to finite physical volume and the fixed topological
charge in our simulation \cite{Aoki:2007ka} are unimportant for the
short-distance quantities considered in this work. 
A simple order counting gives an error of order 
$1/(QL)^2\lesssim$ 0.4\% or smaller. 

To quote the final result, we take the central value from the fit
without the discretization effect 
\begin{equation}
  \Lambda^{(2)}_{\overline{MS}} = 0.234(9)(^{+16}_{-\ 0}) \mathrm{~GeV},
\end{equation}
where the first error is statistical and the second is systematic
due to the discretization and truncation errors.
The result is compatible with previous calculations of $\alpha_s$ in 
two-flavor QCD: 
$\Lambda^{(2)}_{\overline{MS}}$ = 
0.250(16)(16)~GeV \cite{DellaMorte:2004bc} and 
0.249(16)(25)~GeV \cite{Gockeler:2005rv}.
(The physical scale is normalized with an input $r_0$ = 0.49~fm.)

\subsection{$V-A$ channel}

\begin{figure}
  \begin{center}
    \includegraphics[width=100mm]{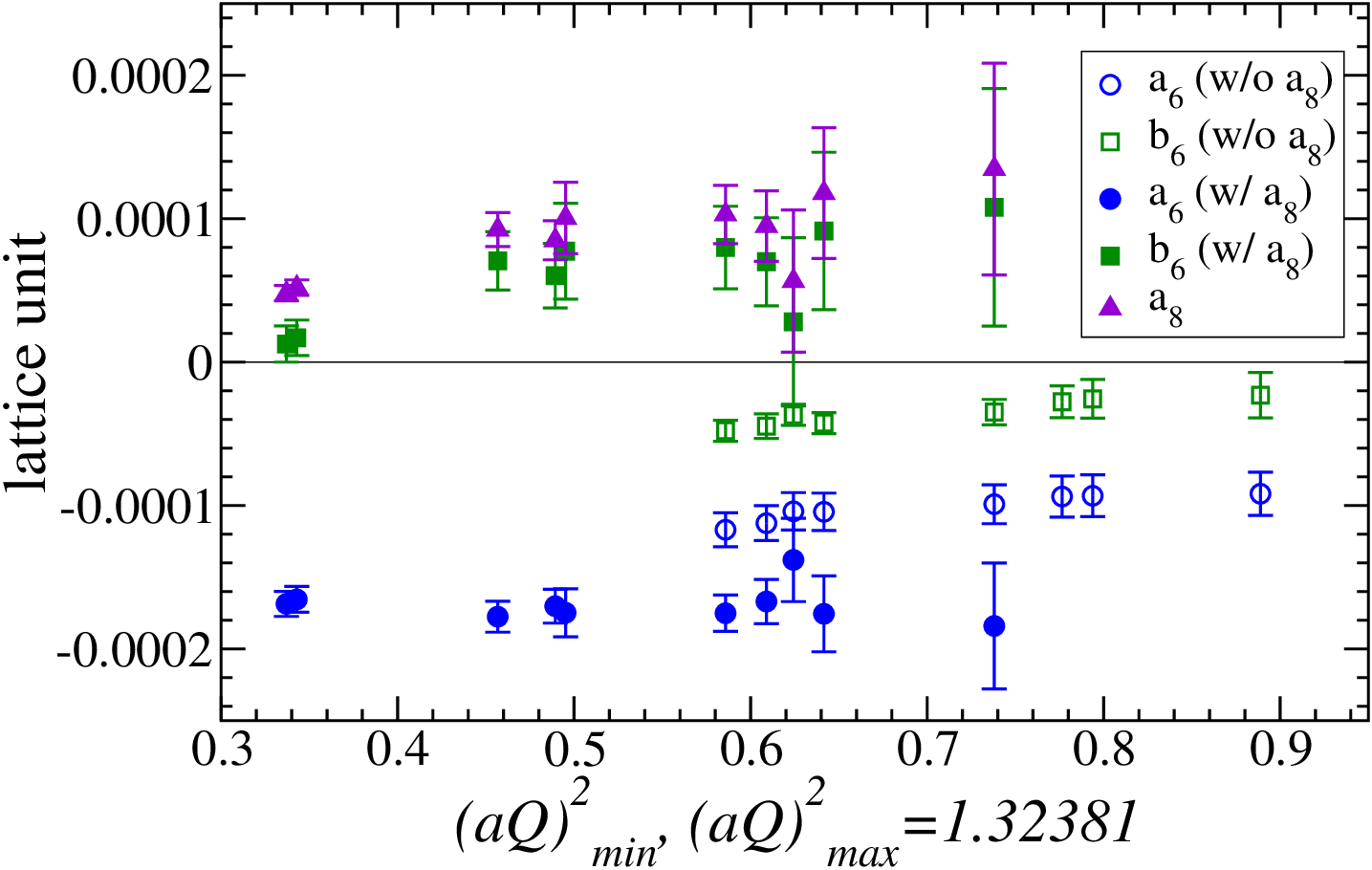}
    \caption{
      Fit range dependence of $a_6(\mu)$, $b_6(\mu)$ and $a_8$.
      The horizontal axis denotes the minimum momentum squared
      $(aQ)_{\rm min}^2$.
    }
    \label{fig5a}
  \end{center}
\end{figure}

For the fit of the $V-A$ vacuum polarization $\Pi_{V-A}^{(0+1)}(Q)$,
we also examine the fit range dependence.
In Figure~\ref{fig5a} the fit parameters $a_6$, $b_6$ and $a_8$ are
shown as a function of $(aQ)^2_{\mathrm{min}}$ while fixing
$(aQ)^2_{\mathrm{max}}$ at the same value 1.3238.
We attempt to fit with (filled symbols) and without (open symbols) the
$a_8/Q^8$ term in order to investigate how stable the results are
against the change of the order of the $1/Q^2$ expansion.
We find that the fit with $a_8/Q^8$ is stable down to 
$(aQ)_{\mathrm{min}}^2\simeq 0.46$, while the other could not be
extended below $(aQ)_{\mathrm{min}}^2\simeq 0.58$.
The difference between filled and open symbols is marginal for $a_6$
(circles), but too large to make a reliable prediction for $b_6$ (squares). 
To quote the results we set $(aQ)_{\mathrm{min}}^2=0.586$ for both
$\Pi^{(0+1)}_{V+A}(Q)$ and $\Pi^{(0+1)}_{V-A}(Q)$.

\begin{figure}
  \begin{center}
    \includegraphics[width=100mm]{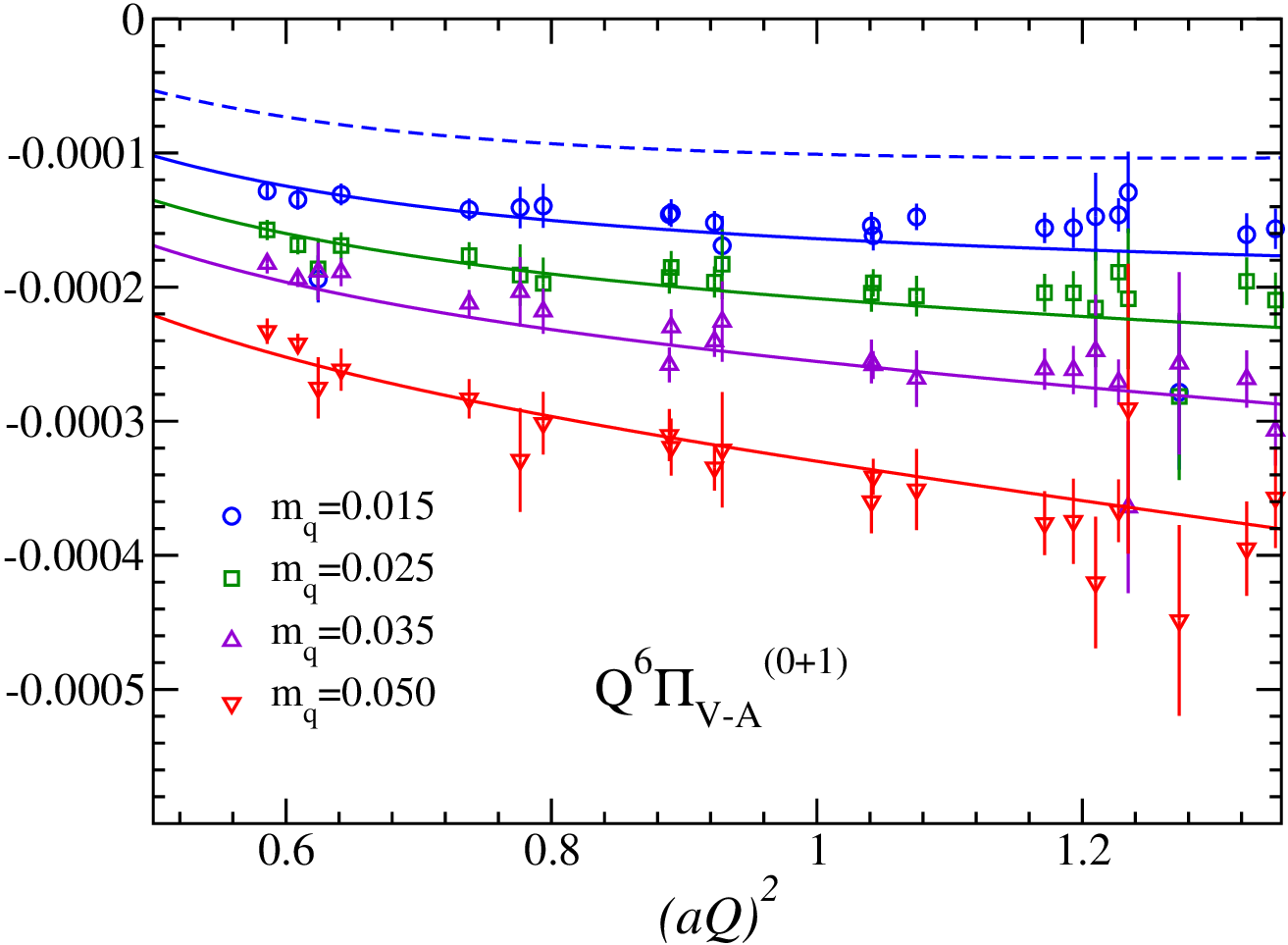}
    \caption{
      $Q^6\Pi_{V-A}^{(0+1)}(Q)$ as a function of $(aQ)^2$.
      The lattice data at different quark masses are shown by open symbols.
      Fit curves for each quark mass and in the chiral limit are drawn. 
    }
    \label{fig5b}
  \end{center}
\end{figure}

In Figure~\ref{fig5b}, we plot $Q^6\Pi_{V-A}^{(0+1)}(Q)$ as a function
of $(aQ)^2$ for four different values of the quark mass $m_q$.
The quark mass dependence is clearly observed.
The main contribution comes from a dimension-six term $c_6 m_q/Q^6$, while the
dimension-four term $\langle m\bar qq\rangle/Q^4$ is sub-dominant
($\sim 20\%$), as its coefficient starts at $\mathcal O(\alpha_s)$.
In the chiral limit, there is a small but non-zero value remaining in 
$Q^6\Pi_{V-A}^{(0+1)}|_{\rm OPE}(Q^2)$ as shown by a dashed curve in the plot.
This is due to the four-quark condensates $a_6$ and $b_6$.

The four-quark condensate $a_6$ obtained from $\Pi_{V-A}^{(0+1)}(Q)$ is
\begin{equation}
  a_6(2\mbox{~GeV}) = -0.0038(3)(^{+16}_{-\ 0})\,{\rm GeV^6},
\end{equation}
where the first error is statistical.
The second error represents an uncertainty due to the truncation of
the $1/Q^2$ expansion.
The central value is taken from the fit with $a_8/Q^8$ in
(\ref{eq:pi_V-A_OPE}) and the error reflects the shift when this term
is discarded.
The result agrees with the previous phenomenological estimates
$-(0.003\sim 0.009)$~GeV$^6$ \cite{Almasy:2008al}.
The other condensate is less stable; we obtain
$b_6(2~GeV)$ = $+$0.0017(7)~GeV$^6$ or $-$0.0008(2)~GeV$^6$
with or without the $\mathcal{O}(1/Q^8)$ term, respectively.

\section{Conclusion}
\label{sec:conclusion}

Many of the lattice calculations to date have analyzed the two-point
correlation functions to extract physical quantities such as the
hadron mass spectra and decay constants.
Usually the exponential fall-off of the correlator at large Euclidean
time separation is used to isolate the ground state contribution.
In this way, however, many interesting pieces of information are
lost.
They are in the short and middle distance regime where the
perturbative analysis is also applicable.
We use the two-point current correlators calculated on the lattice to
extract the strong coupling constant with the help of the continuum
perturbation theory and the operator product expansion.
The recent work by Allison {\it et al.} has exploited
\cite{Allison:2008xk} the similar idea and applied it to the charmonium
correlator to extract the charm quark mass and the strong coupling
constant. 

With the exact chiral symmetry realized by the overlap fermion
formulation, the analysis of the lattice data is simplified. 
For the case of the vacuum polarizations, the continuum form of OPE
may be applied without suffering from additional operator mixings,
such as the additive renormalization of the operator $\bar{q}q$, which
appears in the Wilson-type fermion formulations.
We also obtain the four-quark condensates $a_6$ and
$b_6$, which are relevant to the analysis of kaon decays \cite{Donoghue:1999ku}.

In principle, our analysis does not require lattice perturbation
theory, which is too complicated to carry out to the loop orders
available in the continuum theory.
But the perturbative calculation is still useful to estimate the
discretization effects, which is well-described by perturbation
theory in the asymptotic free theories.

The result for the strong coupling constant is compatible with
previous lattice calculations.
The size of statistical and systematic errors is also comparable with
them. 
An obvious extension of this work is the calculation in 2+1-flavor
QCD, which is underway \cite{Hashimoto:2007vv}. 
We also study the improvement of the analysis by using the conserved
current for the overlap fermion formulation.

\begin{acknowledgments}
Numerical calculations are performed on IBM System Blue Gene Solution
and Hitachi SR11000 
at High Energy Accelerator Research Organization (KEK) 
under a support of its Large Scale Simulation Program (No.~07-16).
This work is supported by the Grant-in-Aid of the Japanese Ministry of Education
(No. 18034011, 
     18340075, 
     18740167, 
     19540286, 
     19740121, 
     19740160, 
     20025010, 
     20340047, 
     20740156),  
and National Science Council of Taiwan (No.~NSC96-2112-M-002-020-MY3, NSC96-2112-M-001-017-MY3).
\end{acknowledgments}

\end{document}